\documentclass[review]{elsarticle}

\usepackage{lineno,hyperref}
\modulolinenumbers[5]

\begin{document}

\begin{frontmatter}

\title{Isotope production in thunderstorms}

\author{Pablo G. Ortega}
\address{Departamento de F\'isica Fundamental and Instituto Universitario de F\'isica Fundamental y Matem\'aticas (IUFFyM), Universidad de Salamanca, E-37008 Salamanca, Spain}
\ead{pgortega@usal.es}

\begin{abstract}
Thunderstorms and lightnings are natural particle accelerator systems. Terrestrial gamma-ray flashes, caused by relativistic runaway electron avalanches, produce bursts of $X$ and $\gamma$ rays, energetic enough to produce photo-nuclear reactions within the atmosphere. Such reactions cause the generation of new isotopes, which modify the air composition locally and produce new ways of detecting and characterizing this high-energy phenomena.
In this work we explore, using the general purpose Monte Carlo transport code FLUKA, the production of secondaries after a Terrestrial gamma-ray flash and analyze the generation of new isotopes in detail. Their abundance, time and energy profiles are studied, which can be useful for establishing new measuring strategies.
\end{abstract}

\begin{keyword}
Thunderstorms\sep Lightning\sep Atmospheric electricity\sep Terrestrial Gamma-ray flash\sep Isotopes
\MSC[2020] 86A10\sep  76X05\sep 81V35\sep 82D75
\end{keyword}

\end{frontmatter}


\section{Introduction}
Earth's atmosphere is constantly flooded with energetic radiation of diverse origins. 
The most abundant and widely explored, since Victor Hess made his pioneer studies on the rate of 
ionisation in the atmosphere in 1912~\cite{Hess:1912srp,hess2018observations}, is the radiation coming from extraterrestrial sources, mostly from
high energetic protons and charged nuclei from the solar wind or distant stars and galaxies.
However, in the last decades a new source of high-energy particles has been identified, this time produced 
within the atmosphere: that associated with thunderclouds and lightnings. 

Traditionally, the electrical discharges and cloud electrification have been investigated 
by means of classical Electromagnetism, due to the usual energy ranges involved in such phenomena (in the order of units and tens of eV). 
The situation changed in the 60's, when numerous studies observed the emission of X-radiation and runaway electrons in the nanosecond-scale from gas discharges in the laboratory, at pressures from tens to thousands of Torr and high-voltage pulses\cite{1968SPhD...12.1042S,1970SPTP...14.1148T,Tarasova1974FastEA,Babich_1990}. Later on, in 1994, Fishman {\it et al} observed 
brief, intense flashes of gamma rays in orbiting detectors, 
which were originated in the atmosphere~\cite{Fishman1313}.
Those brief bursts of $\gamma$-rays, dubbed as Terrestrial Gamma-ray Flashes (TGFs), 
were found to be produced in coincidence with lightning discharges, and their energy spectra 
was consistent with bremsstrahlung emissions from MeV electrons.
Since then, similar events compatible with TGFs related to lightnings have been observed in $\gamma$-ray satellite detectors~\cite{Smith1085,doi:10.1029/2009JA014502,doi:10.1029/2019JD031214,doi:10.1029/2017JA024837,URSI201750},  
airborne observations~\cite{doi:10.1029/2011JD016252,doi:10.1029/2017JD027771}, in-flight 511 keV enhancement observations hypothesized to be TGF residuals~\cite{dwyer_smith2015,doi:10.1029/2018JD028337} and 
ground-based detectors~\cite{brunetti2000gamma,chubenko2000intensive,torii2002observation,torii2004downward,tsuchiya2007detection,tsuchiya2009observation,PhysRevLett.123.061103,doi:10.1029/2017JD027931,doi:10.1002/2017GL075071} (though for some of these works the events are most likely gamma glows related to lightnings).

The emission of high energetic photons (X and $\gamma$-rays) is produced over wide time scales, ranging from $X$-rays bursts linked with lightning leaders in the sub-$\mu$s scale to the sub-millisecond scale of $\gamma$-rays (TGFs), followed by the glows observed at thunderstorms in the second to minute scale~\cite{dwyer2012high}. 
Regarding their energy spectra, it is found to be harder than those of other cosmic sources~\cite{dwyer2012high}. 
The $\gamma$'s from TGFs observed by the RHESSI~\footnote{Reuven Ramaty High Energy Solar Spectroscopic Imager} spacecraft, a NASA's satellite detector designed to study $X$ and $\gamma$-rays from solar flares, have energies up to 20 MeV. 
One of the most popular candidates
to explain the TGF is the generation of relativistic runaway electron avalanches (RREAs)~\cite{GUREVICH1992463,BABICH1998460,Babich2004exp,doi:10.1029/2004GL019795}. 
Such avalanche of relativistic electrons, produced via ionization collisions~\cite{GUREVICH1992463,doi:10.1029/1999JA900335,doi:10.1029/2004GL019795}, 
occurs for electric fields above $E_{\rm th}=284$ kV/m $\times (\rho/\rho_0)$, where $\rho$ is the density of air and $\rho_0$ the density of air at STP~\footnote{Standard Temperature and Pressure}.
However, pure RREA mechanism is not capable of predicting the observed TGFs intensities. Indeed, basing on the available experimental data, there is still not a clear scenario in which RREA, acting on external $e^-$ population, is the unique mechanism for generating TGFs~\cite{doi:10.1029/2007JD009248}. Further mechanisms, such as relativistic feedback, mainly of photons and positrons that create new avalanches~\cite{doi:10.1029/2004GL021744,doi:10.1029/2011JA017160}, are needed to properly describe TGF fluxes.

The emission of bremsstrahlung $\gamma$ from the initial RREA follows, approximately, a $E^{-1}e^{-E/E_0}$ distribution, with $E_0=7.3$ MeV~\cite{dwyer2012high}. The ensuing interactions caused by the high-energy gammas are dominated by electromagnetic interactions, producing secondary avalanches of $e^\pm$ and lower energy $\gamma$'s. 
However, if the energy of the TGF's $\gamma$-rays is sufficiently large, they can trigger atmospheric photo-nuclear reactions and generate fast neutrons~\cite{10.1029/2006JD008340,Babich2006,Babich2007}. The most common photo-nuclear reactions in the atmosphere after a TGF are $\gamma+{^{14}N}\to n+{^{13}N}$, $\gamma+{^{16}O}\to n+{^{15}O}$ and $\gamma+{^{40}Ar}\to n+{^{39}Ar}$, with threshold energies of 10.55 MeV, 15.7 MeV and 9.55 MeV~\cite{Babich:2014bia}, respectively. Considering that the average energy of electrons in a TGF is around $7$ MeV~\cite{dwyer2012high}, we observe that the production of neutrons by these mechanisms is possible and, actually, quite efficient. Indeed, it is estimated that around $10^{12}$ to $10^{15}$ neutrons can be produced by a typical TGF, depending on the initial assumptions taken~\cite{Babich_2019,doi:10.1002/2017GL075552,doi:10.1029/2009JA014696}. 

Furthermore, these photo-nuclear reactions have another consequence: the generation of new isotopes, which can  modify locally the atmosphere's composition. 
The aforementioned atmospheric photo-nuclear reactions $\gamma+{^{14}N}\to n+{^{13}N}$ and $\gamma+{^{16}O}\to n+{^{15}O}$ generate fast neutrons with kinetic energies around $10$ MeV and further unstable $\beta^+$ emitters, which generate positrons~\cite{Babich_2019,doi:10.1029/JB078i026p05902,doi:10.1002/2017GL075131,Enoto2017,babich2017thunderous}. Those fast neutrons are ultimately stopped via multiple elastic scattering with air nuclei down to thermal energies, being most of them absorbed by nitrogen atoms via the reaction $^{14}N+n\to ^{14}C+{^1H}$. The short-lived $\beta^+$ isotopes $^{13}$N and $^{15}$O, and others, can be observed by measuring the $0.511$ MeV line emission glow in the timescale of seconds to minutes~\cite{torii2004downward,tsuchiya2007detection,tsuchiya2009observation,PhysRevLett.123.061103,Enoto2017,Enoto:2017lpx}, which provides a new method to characterize TGFs via the detection of annihilation gamma intensity and their time profile. The remaining small fraction of stable isotopes, such as $^{13}C$, $^{14}C$ and $^{15}N$, are incorporated to the natural isotope composition on Earth. This is interesting as the possible local enhancement of $^{14}C$ isotopes may have an impact on the use of this isotope as dating method~\cite{10.1029/2006JD008340,Babich_2019,doi:10.1029/JB078i026p05902,doi:10.1002/2017GL075131,babich2017thunderous,Enoto:2017lpx}.
In Ref.~\cite{doi:10.1002/2017GL075131}, indeed, L.~P.~Babich explores the production of radiocarbon $^{14}C$ in thunderstorms by estimating the required neutron fluence per thunderstorm so the $^{14}C$ production is comparable with the cosmic sources, finding that thunderstorm-produced neutrons may locally contribute to the $^{14}C$ concentration, specially in tropics where the thunderstorm activity is especially severe.

Although the main mechanisms for isotope production via photo-nuclear reactions are known, a thorough study of the isotope generation in a typical TGF event has not yet been performed.
In this work we explore in detail the generation of secondaries in a TGF after a runaway electron avalanche, and analyze the production of new isotopes as a consequence of photo-nuclear reactions.

\section{Materials and Methods}

\subsection{Simulation Setup}\label{sec:simsetup}

In this study the abundance of those isotopes and the detection potential of unstable nuclei are explored using the general use Monte Carlo transport code FLUKA~\cite{Ferrari:2005zk,Bohlen:2014buj,Battistoni:2015epi}.
FLUKA is a general purpose Monte Carlo particle transport code which describes the 
interaction of particles and the generation of secondary particles whilst 
traveling through matter. Due to the broad range of energies the FLUKA code is able to manage, from the MeV scale up to TeV,
it has many applications in fields such as high energy experimental physics, shielding, 
detector and telescope design, cosmic ray studies, dosimetry, medical physics and radio-biology. 
The code is built and maintained with the purpose of providing the most complete and 
precise possible physical models. 
The implementation of accurate nuclear and electromagnetic models includes the relevant 
processes to perform studies for high energy atmospheric events such as Terrestrial Gamma-Ray Flashes (see for example Ref.~\cite{Rutjes:2016a,Rutjes:2017a}). 
Furthermore, the code has recently included model refinements in nuclear 
interactions and prompt-gamma generation~\cite{Bohlen:2014buj}, which reproduce experimental measurements within $10-15\%$. The code has also been employed for cosmic ray events simulations, specifically to evaluate the atmospheric neutrino flux in a 3D model of the atmosphere of the Earth~\cite{Battistoni:1999at,Battistoni:2002ew}, exploiting the code's high degree accuracy in the
description of hadron-hadron interactions and particles production.

In this work the FLUKA version 2011-2x.8 was used. All the relevant physics is included in the code by default. Transport options were selected to enable electromagnetic showers, Rayleigh scattering and inelastic 
form factor corrections for Compton scatterings, detailed photoelectric edge 
treatment and fluorescence photons, fully analogue absorption for low-energy 
neutrons and restricted ionization fluctuations on both hadrons and leptons. Thermal neutrons were 
transported down to $10^{-5}$ eV; other particles were transported down to 100 keV. Delta-ray production threshold was set to 100 keV. To calculate energy loss by hadrons and muons, momentum loss per unit distance ($dp/dx$) tabulations are used, with 80 logarithmic intervals or logarithmic widths with a ratio of $1.04$ between the upper and the lower limit of each interval, whichever is more accurate. The fractional kinetic energy loss per step was set to 5\%. Also, heavy particle $e^+/e^-$ pair production and 
bremsstrahlung are activated, with full explicit production and transport of secondary particles, 
with the minimum threshold equal to twice the electron mass for pair 
production and photon production above $300\,keV$ for bremsstrahlung. 
Photo-nuclear interaction is activated at all energies, with explicit generation of secondaries and
heavy fragment transport is, as well, activated.
In order to obtain accurate results for residual nuclei and fragment production we have additionally activated coalescence mechanisms, ion splitting into nucleons, ion electromagnetic-dissociation mechanisms and evaporation models of heavy fragments.

The creation of isotopes from photo-nuclear reactions originated after a
Terrestrial gamma-ray flash is evaluated starting from a typical spectra of relativistic runaway electron avalanche in strong electric fields of lightning discharges~\cite{doi:10.1029/2011JA016494,PhysRevLett.123.061103,dwyer2012high}. Primary particles are injected at three different heights from the mean sea level: $2.5$ km, $5$ km and $10$ km, covering the usual vertical extend of 
this type of events. Those primary electrons are emitted isotropically from a point-like source at each height. 
No broadening of the source is considered, nor it has an expected impact on the results given the large natural spread of secondaries~\footnote{Note that the usual extend of TGF is much lower than the atmospheric mean free path. }. Besides, no geomagnetic model has been considered. 

Several simulations of runaway electron avalanches show a similar average energy for the initial $e^-$ population, of $7.3$ MeV~\cite{dwyer2012high}, quite independent of the electric field strength over a wide range. Then, one would expect an exponentially decaying energy spectrum
valid up to the maximum kinetic energy of the runaway electrons, given by the properties of the electric field strength of the thunderstorm discharge. Thus, the energy of the
electrons is randomly sampled from an exponential probability distribution function, $e^{-E/E_0}$, with $E_0=7.3$ MeV following the predicted spectrum of a relativistic runaway electron avalanche~\cite{doi:10.1029/2011JA016494,PhysRevLett.123.061103,dwyer2012high}, choosing a minimum energy of $1$ MeV.

A 3D cylindrical modelling of the Earth's atmosphere has been implemented for the simulation, covering heights from the ground up to 15 km and an horizontal extend of 10 km radius. This geometry is filled by a medium composed by dry air, given by a proper mixture of Nitrogen, Oxygen, Carbon and Argon (see Table~\ref{tab:dryair} for the precise composition), arranged in successive cylindrical air layers of $0.25$ km height, each one having a density scaling according to the widely-used U.S. Standard atmosphere profile~\cite{united1976u}.
The density effects in the medium are simulated by means of the Sternheimer parameters~\cite{Sternheimer:1971zz,Seltzer:1982ks} summarized in Tab.~\ref{tab:dryair}.

 \begin{table}
 \caption{\label{tab:dryair} Composition of dry air by mass fraction used in this study, and Sternheimer parameters~\cite{Sternheimer:1971zz,Seltzer:1982ks} for the medium.}
 \centering
 \begin{tabular}{l c}
 \hline
  Composition  & Mass fraction  [\%] \\
 \hline
   Carbon  & $0.01248$   \\
   Nitrogen  & $75.5267$  \\
   Oxygen & $23.1781$   \\
   Argon  & $1.2827$   \\
 \hline
 \end{tabular}
 \hfil
 \begin{tabular}{l c}
 \hline
 \multicolumn{2}{l}{Sternheimer parameters:} \\
 \hline
Ionization Potential & $85.7$ eV \\
$\bar C$  & $10.5961$ \\
$X_0$ &  $1.7418$ \\
$X_1$ & $4.2759$ \\
$a$ & $0.10914$ \\
$m$  & $3.3994$ \\
\hline \end{tabular}
 \end{table}

\subsubsection{Scoring Method}

The relevant secondary particles causing nuclear fragmentation in air molecules are photons and neutrons. After the initial RREA, these high-energy electrons are abruptly stopped, generating bremsstrahlung photons. If the energy of 
the latter photons is above the photo-nuclear reaction threshold of about $10$ MeV~\cite{10.1029/2006JD008340}, they will interact with stable air nuclei breaking them apart and emitting neutrons, protons and other fragments. 

Of course, the dominant processes that occur in the simulation are electromagnetic interactions, in the form of $e^\pm$ and secondary $\gamma$ showers. As photo-nuclear cross sections are much smaller than $\gamma$ cross sections for EM interactions with electrons and atoms, the analogue simulation of photo-nuclear interactions is very inefficient. Then, if we want a valuable analysis of ion generation from photo-nuclear interactions we need whether to acquire large statistics (with a heavy CPU time and consumption) or artificially increase the frequency of such inelastic interactions through some adequate bias in the code. In the present study, the photo-nuclear interaction length is reduced by a factor $40$. 
The secondary particles, created from interaction points sampled according to the biased probability, are created with a reduced weight, adjusted to take into account the ratio between biased and physical survival probability.

Two types of scoring methods are implemented. On the one hand, the properties of the new emerging particles is scored at their point of creation. The relevant information includes the type of created particle, its parent particle, the interaction that caused its appearance, the kinetic energy of the particle, its position in space, the time of creation~\footnote{The origin of time is set at the moment the primary particle which caused the chain of interactions that leads to the aforementioned particle's creation started to be transported.} and statistical weight (which encodes any possible bias applied to the simulation).
On the other hand, each separation between air layers in the full atmosphere model at FLUKA acts as a perfect detector, so any particle crossing the boundary between two regions with different density will be scored. In this case, the interesting properties are the type of particles, position of the crossing, the time and the kinetic energy.
All this information emerging from the full simulation is condensed in appropriate histograms, which are 
analyzed in the following section.

A total of $10^{10}$ electrons are simulated for each height. Such number is large enough to acquire statistics in a reasonable CPU time. 
The typical $e^-$ population in a TGF is difficult to measure, but it can be guessed from the radiation detected. Assuming a mono-energetic electrons of $35$ MeV, the authors of Ref.~\cite{Smith1085} estimated around $10^{15}$ electrons per flash. This number is several orders of magnitude lower than other studies, such as Ref.~\cite{PhysRevLett.123.061103} which estimated $\sim 10^{19}$ electron avalanche with energies above $1$ MeV at a downward TGF. Similarly, \cite{mailyan2016spectroscopy} analyzed around 50 TGFs and, for source altitudes above 10 km, constrained the average number of electrons with $E>1$ MeV to around 2$\cdot 10^{18}$, ranging from $4\cdot 10^{16}$ to $3\cdot 10^{19}$. 
The study of Ref.\cite{gjesteland2015observation}, on the contrary, estimated the number of $\gamma$'s above 1 MeV to be between $10^{17}$ and $10^{20}$ which, considering that around $30\%$ of the generated photons have energies above $1$ MeV, throws a number of [$3\cdot10^{17}$,$3\cdot 10^{20}$] electrons. 
Thus,  we can take the conservative value of $10^{18\pm2}$ electrons above $1$ MeV. In this case, the results in this work should be multiplied by a factor $10^{8\pm2}$ to estimate the real abundances.
Nevertheless, the results will expressed per primary electron (above 1 MeV) to avoid further assumptions on the
initial electron population.

In order to have a sense of the statistical uncertainty of the simulation, the full $10^{10}$ are divided in $10^4$ bunches  of $10^6$ primaries each, which are executed independently. The results for each bunch are, then, statistically combined, so the average and standard deviation are obtained for each observable. That way we can quantify the statistical fluctuations of the results.

\section{Results} 
\begin{figure}
 \noindent\includegraphics[width=.55\textwidth]{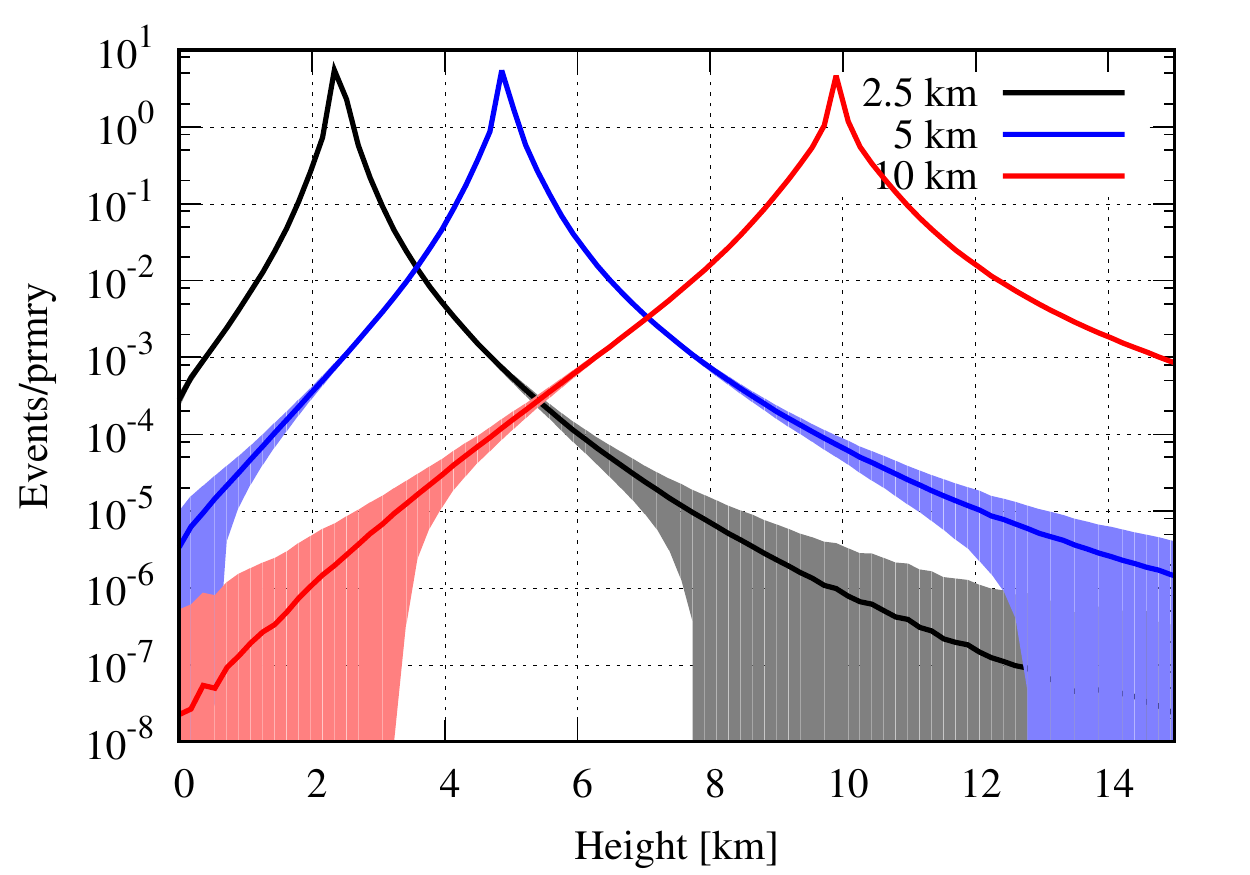}
 \noindent\includegraphics[width=.55\textwidth]{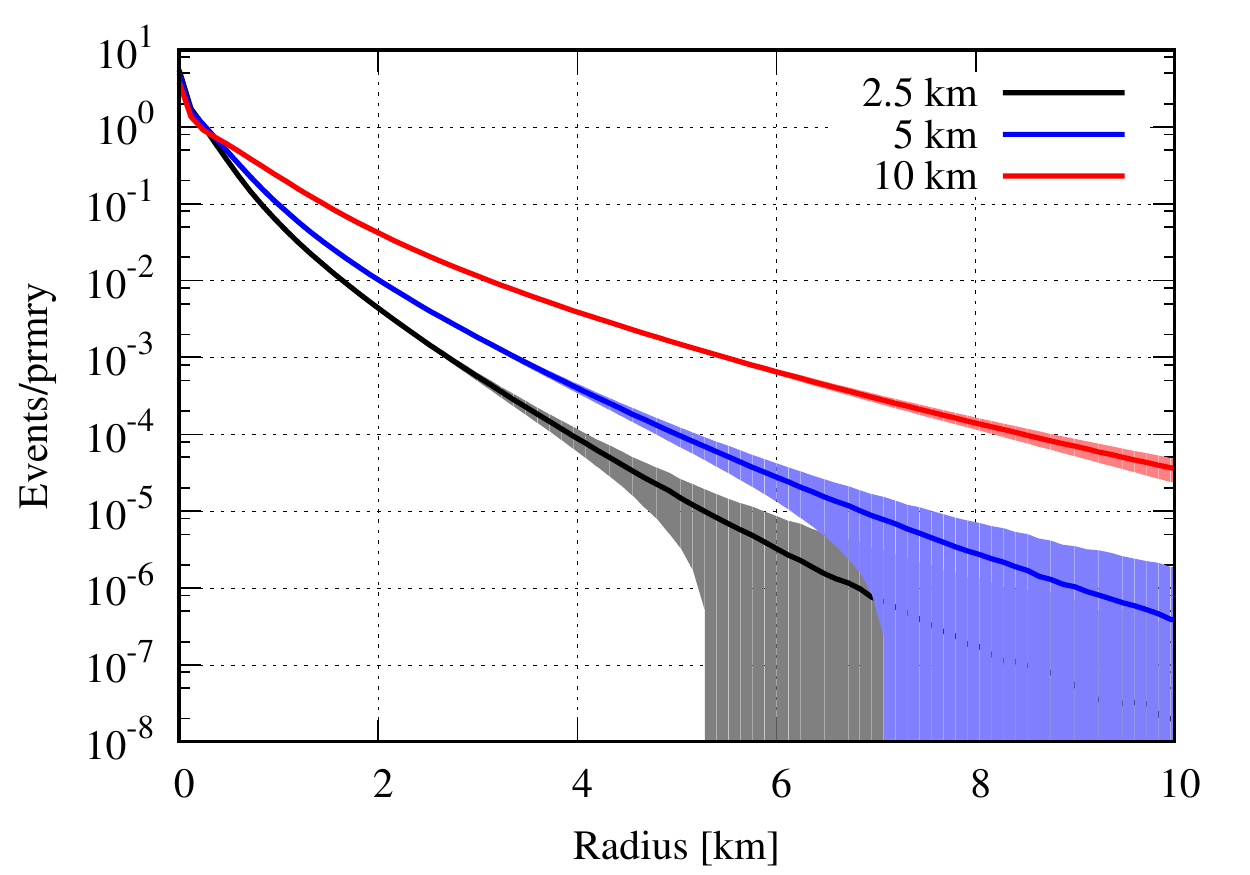}
\caption{Creation position of all secondary particles with the vertical height (left) and the radius (right) for each initial position at $2.5$ km, $5$ km and $10$ km. We observe a larger spread of particles for the $10$ km TGF due to the increase of the atmospheric mean free path as a consequence of the lower density.}
\label{fig:height}
\end{figure}

On a first stage we analyze the secondary creation from the electron runaway avalanche simulation at different heights. In which respects the original height of the RREA, all the secondaries show a similar profile. In Fig.~\ref{fig:height} we show the population of secondaries at different heights, scored at the creation point. The particle range for the higher simulations are larger, given that the density at that points are smaller and, consequently, the mean free path increases. Nevertheless, as the air composition is the same, the secondary spectra are similar. 
Besides that, we observe that, for all cases, most of the secondaries are produced few tens of meters near the source. 

\begin{figure}
\begin{center}
 \noindent\includegraphics[width=.65\textwidth]{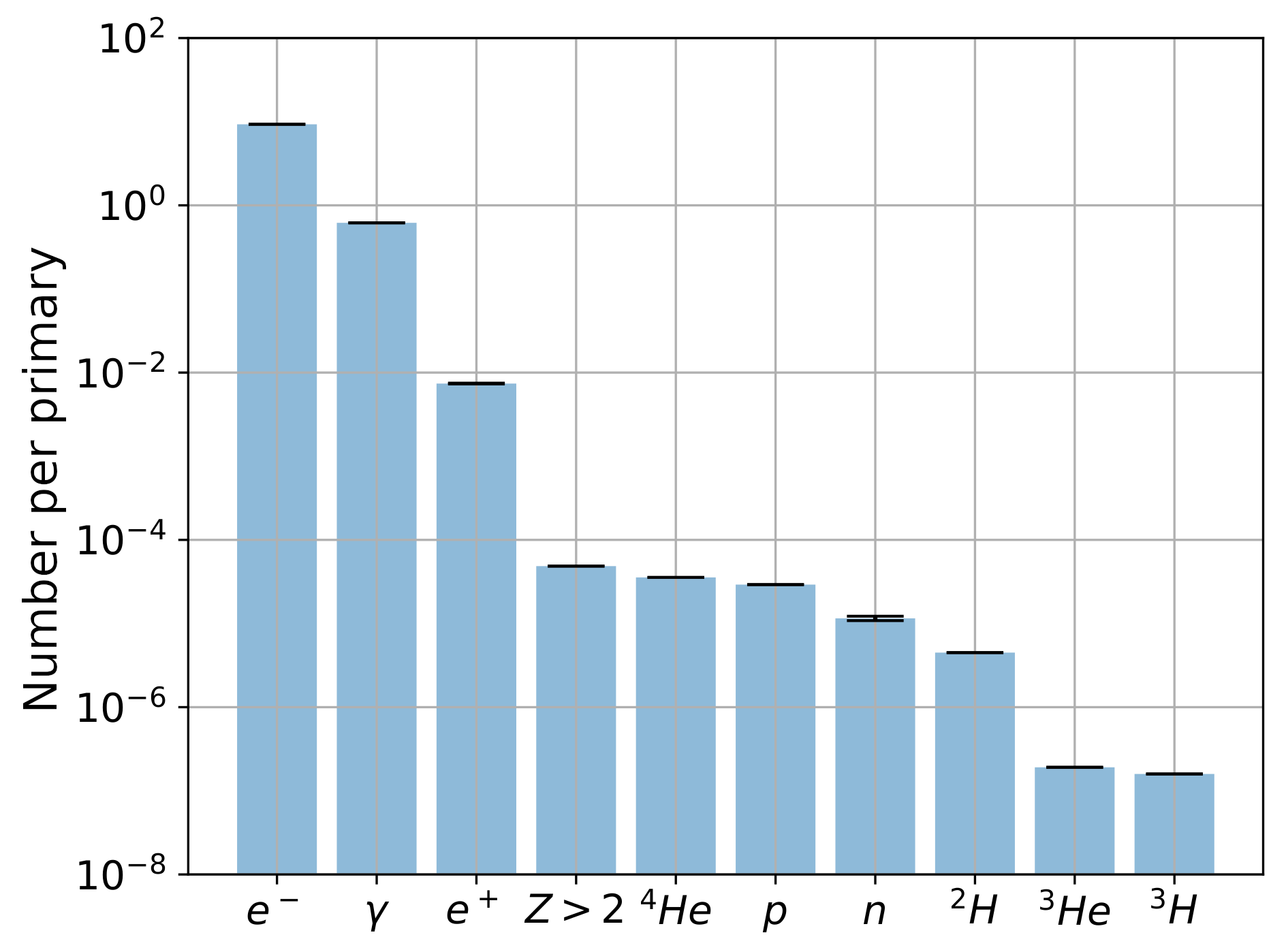}
 \end{center}
\caption{Summary of secondary particles created per primary electron at the TGF event. The $Z>2$ labels all the residual nuclei with atomic number above 2.}
\label{fig:particles}
\end{figure}

A representation of the total population all the created particles for a single RREA event can be seen in Fig.~\ref{fig:particles}. The residual nuclei with $Z>2$ are summed up together under a common label. In order to properly study these residual nuclei their decays have been deactivated in FLUKA, so they are treated as stable for convenience. As expected, the simulation is dominated by EM showers, so $e^\pm$ and $\gamma$ are the most produced secondaries. The positrons, produced by pair production reactions, are rapidly stopped and annihilated with the thermal electrons of the air molecules, producing a large amount of almost-instantaneous $0.511$ MeV annihilation-gamma pairs. These annihilation gammas represents, though, only a $\sim$2.4(1)\% of the total $\gamma$ production~\footnote{We show in parenthesis the 67\% C.L. of the central value.}. The relevant secondaries for isotope creation are the most energetic secondary $\gamma$'s, obtained from bremsstrahlung, which takes a $\sim$97.6(4)\% of the total $\gamma$ production. There is a negligible contribution from inelastic interactions (mainly nuclear de-excitations, which represents a $2.5(5)\cdot 10^{-3}$\%), $\delta$-ray production ($5(1)\cdot 10^{-2}$\%) and low-energy neutron scattering ($3(1)\cdot 10^{-4}\%$). All these $\gamma$'s, which cover an energy range between few MeV and 50 MeV, are responsible of the ion fragmentation that we will analyze later on. EM particles ($e^\pm$ and $\gamma$'s) below that energy represents most of the secondaries population, but are safely stopped and absorbed within the atmosphere. If the RREA is produced at low heights, those EM showers can reach the ground and be measured. Thus, they can give us information on the intensity of the event, provided that the location is known by other means.

Neutrons and protons represent a large portion of the results, similar in production intensity ($2.920(3)\cdot 10^{-5}$ per primary electron for $p$ and $1.15(7)\cdot 10^{-5}$ per primary for $n$). They are the responsible for most isotope generation via spallation reactions. The neutrons, with energies ranging up to $\sim 20$ MeV, are all produced by photo-nuclear inelastic interactions few nanoseconds after the primary electrons starts to be transported in FLUKA. Regarding the protons, their are produced mostly by inelastic interactions ($68\pm9$\% of them), but there is a significant population (of $32\pm8$\%) created from low-energy neutron scattering, with energies between $0$ and $2$ MeV. An example of these kind of $p$ production is the generation of $^{14}C$ that we will evaluate later on. 

\begin{figure}
 \noindent\includegraphics[width=.5\textwidth]{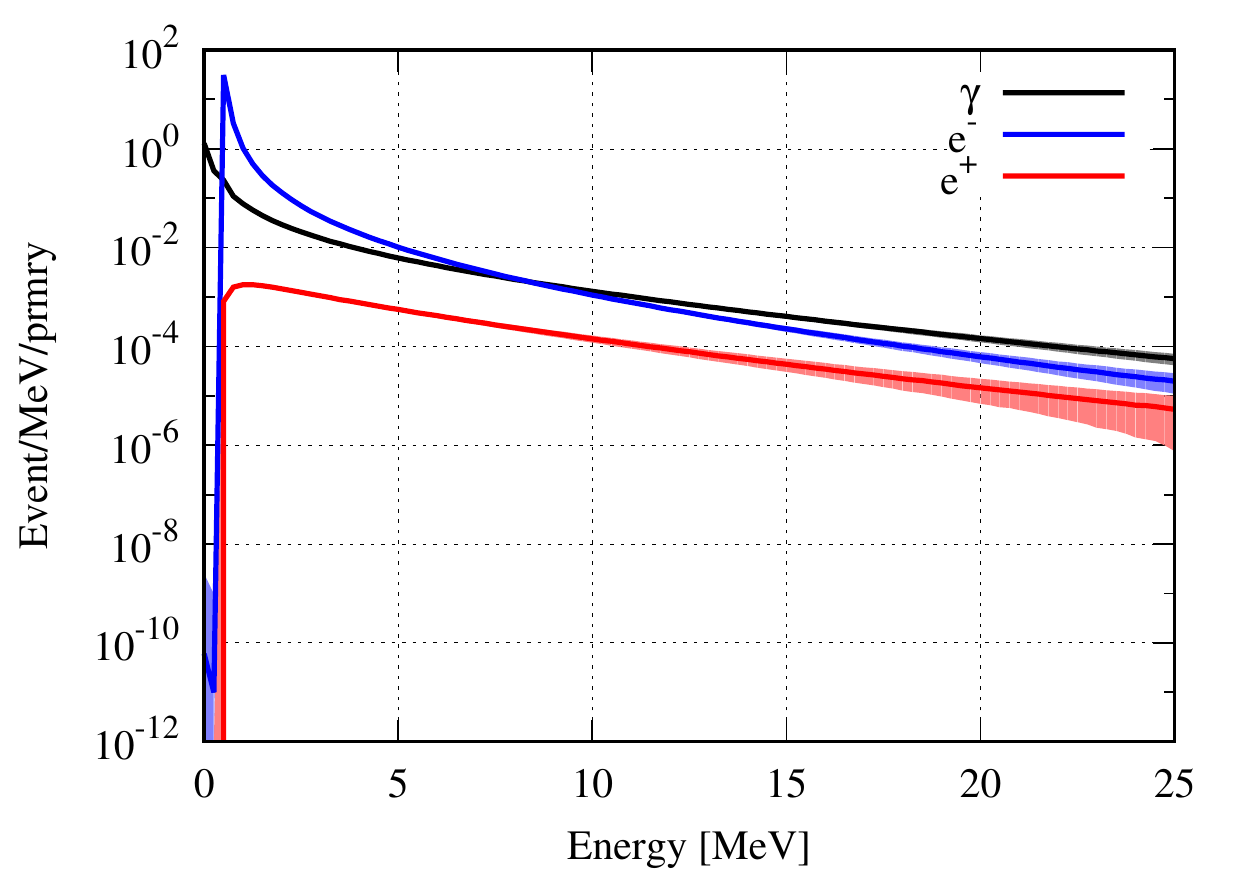}
 \noindent\includegraphics[width=.5\textwidth]{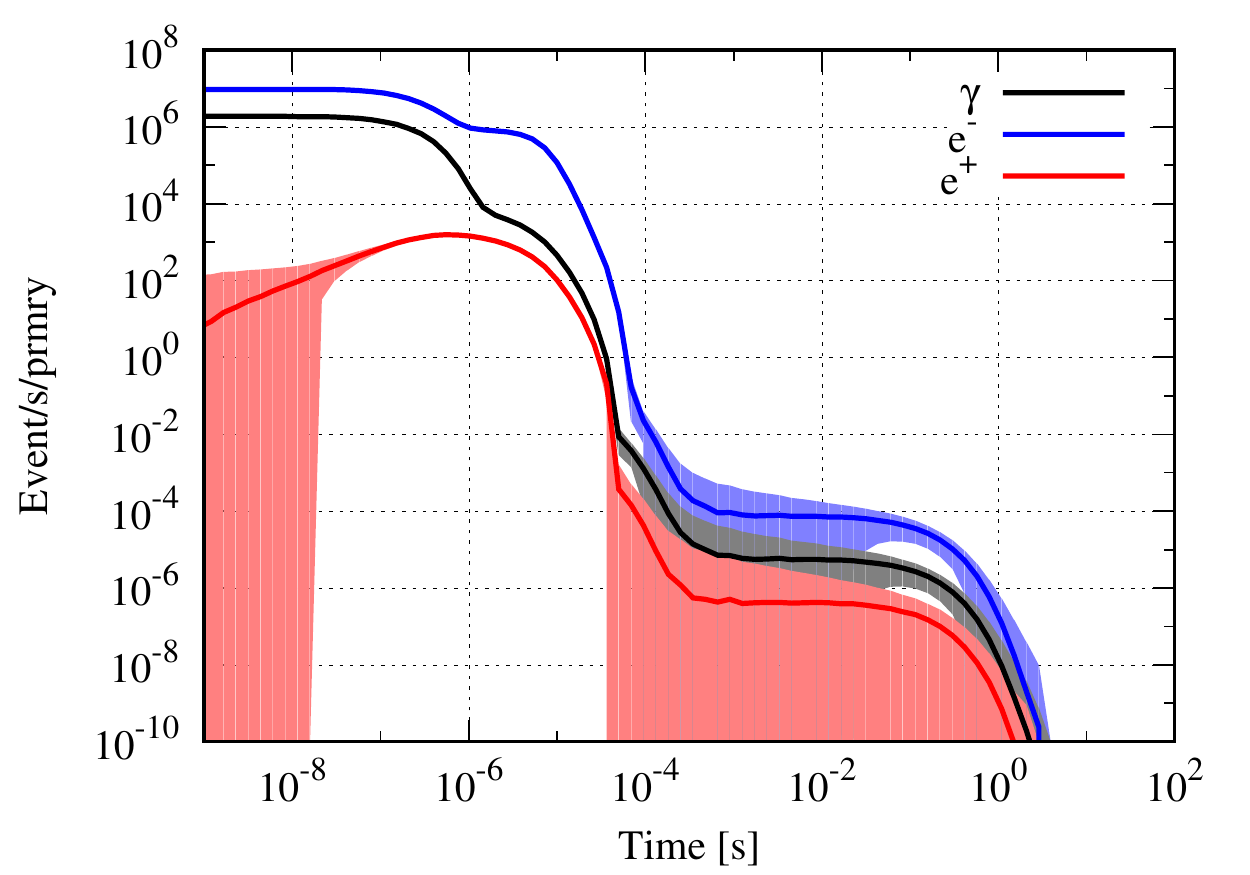}
\caption{Left: Number of created $\gamma$ and $e^\pm$ following the electron runaway avalanche as a function of their energy, shown per primary and energy bin size. Right: Number of created $\gamma$ and $e^\pm$ per primary and time bin size as a function of their time of creation after the primary electrons start to be transported by the code. We observe the main TGF burst few microseconds after and the TGF afterglows in the scale of milliseconds. The $e^\pm$ distributions follow the same trend.}
\label{fig:gammas}
\end{figure}

There is a large amount of light nuclei created by the simulation. The most common ones are $\alpha$ and deuteron particles ($^4He$ and $^2H$ nuclei), which are produced at a rate of around $10^{-5}$ per primary electron ($3.560(4)\cdot10^{-5}$ for $\alpha$'s and $4.490(4)\cdot 10^{-6}$ for $d$). That is translated to an intensity of $10^{13\pm2}$ nuclei considering $10^{18\pm2}$ electrons per TGF. Tritium and $^3He$ are also generated, but at a lower rate: around a factor $20$-$30$ lower than deuteron (see Fig.~\ref{fig:particles}). The rate of heavier isotopes (with $Z\ge 3$) is $4.859(3)\cdot 10^{-5}$ per primary electron, which is considerably high. The main responsible for their creation are $\gamma$'s with energies above $10$ MeV, which represent only a $\sim 1\%$ of the  $\gamma$'s produced in the simulated relativistic runaway electron avalanche. The time and energy profile of all the $\gamma$'s is shown in Fig.~\ref{fig:gammas}, together with $e^{\pm}$ particles. The $\gamma$ spectrum agrees with the expected $E^{-1}e^{-E/7.3{\rm MeV}}$ distribution from bremsstrahlung~\cite{dwyer2012high}.

In order to analyze in more detail the isotope generation from TGFs, we performed a second step simulation which uses $\gamma$ particles above $10$ MeV as primaries, sampled from a $E^{-1}e^{-E/7.3{\rm MeV}}$ energy distribution. Regarding the isotope creation, we have checked that the use of $\gamma$'s over $10$ MeV gives the same results as those given by the simulation with electrons. The advantage is, of course, the large statistical gain of the $\gamma$ simulation. The lower energy cut is set to $10$ MeV, close to the lower photo-nuclear interaction threshold. As $\gamma$'s above $10$ MeV represent only a $1\%$ of the total $\gamma$'s produced in a the simulated runaway electron avalanche, a correction factor of $10^{-3}$ is applied to the following results, so they are still normalized per primary electron.
For this second simulation, a total number of $2\cdot 10^9$ $\gamma$ primaries have been simulated.

 \begin{figure}
 \noindent\includegraphics[width=\textwidth]{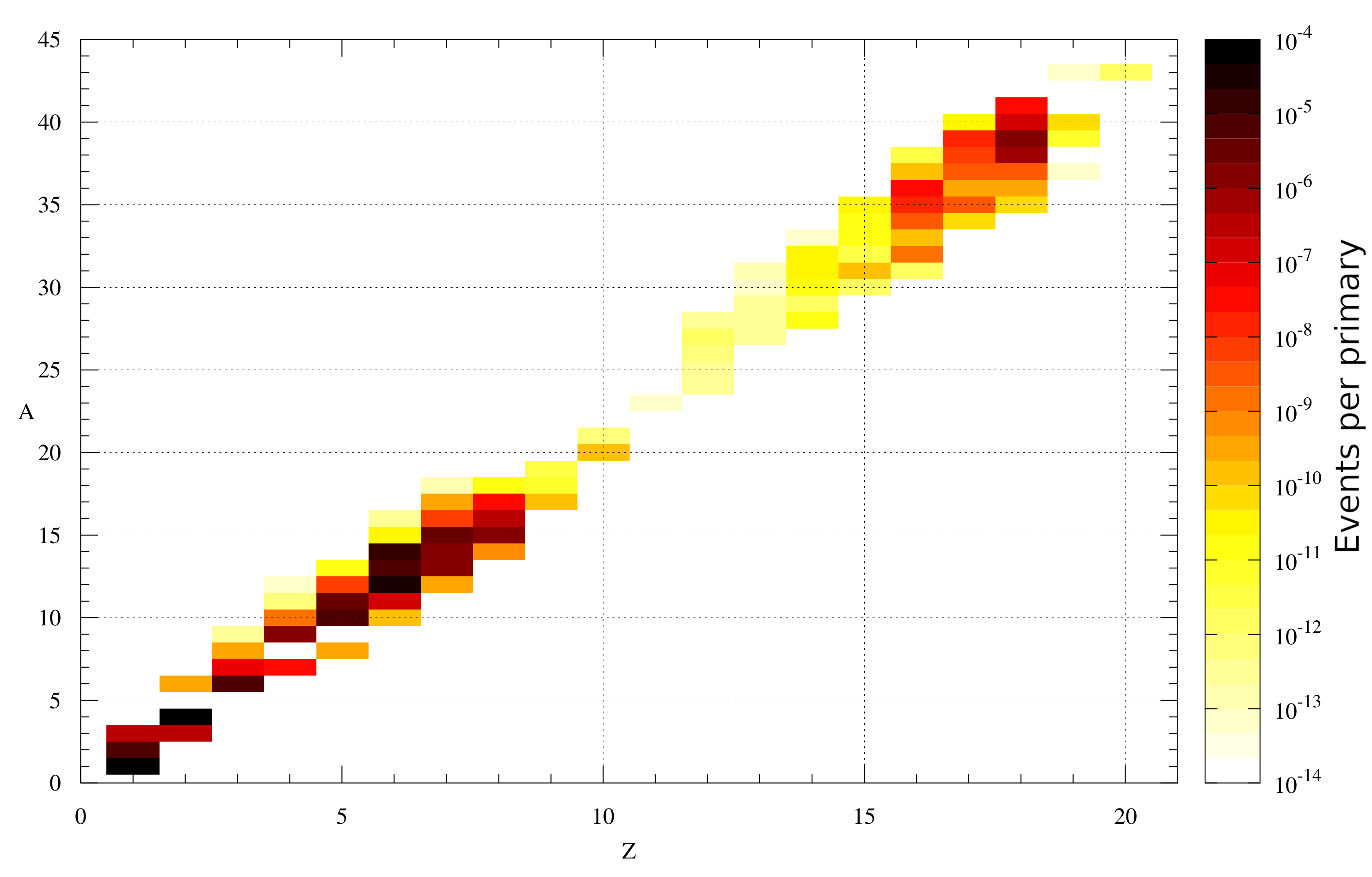}
 \noindent\includegraphics[width=.5\textwidth]{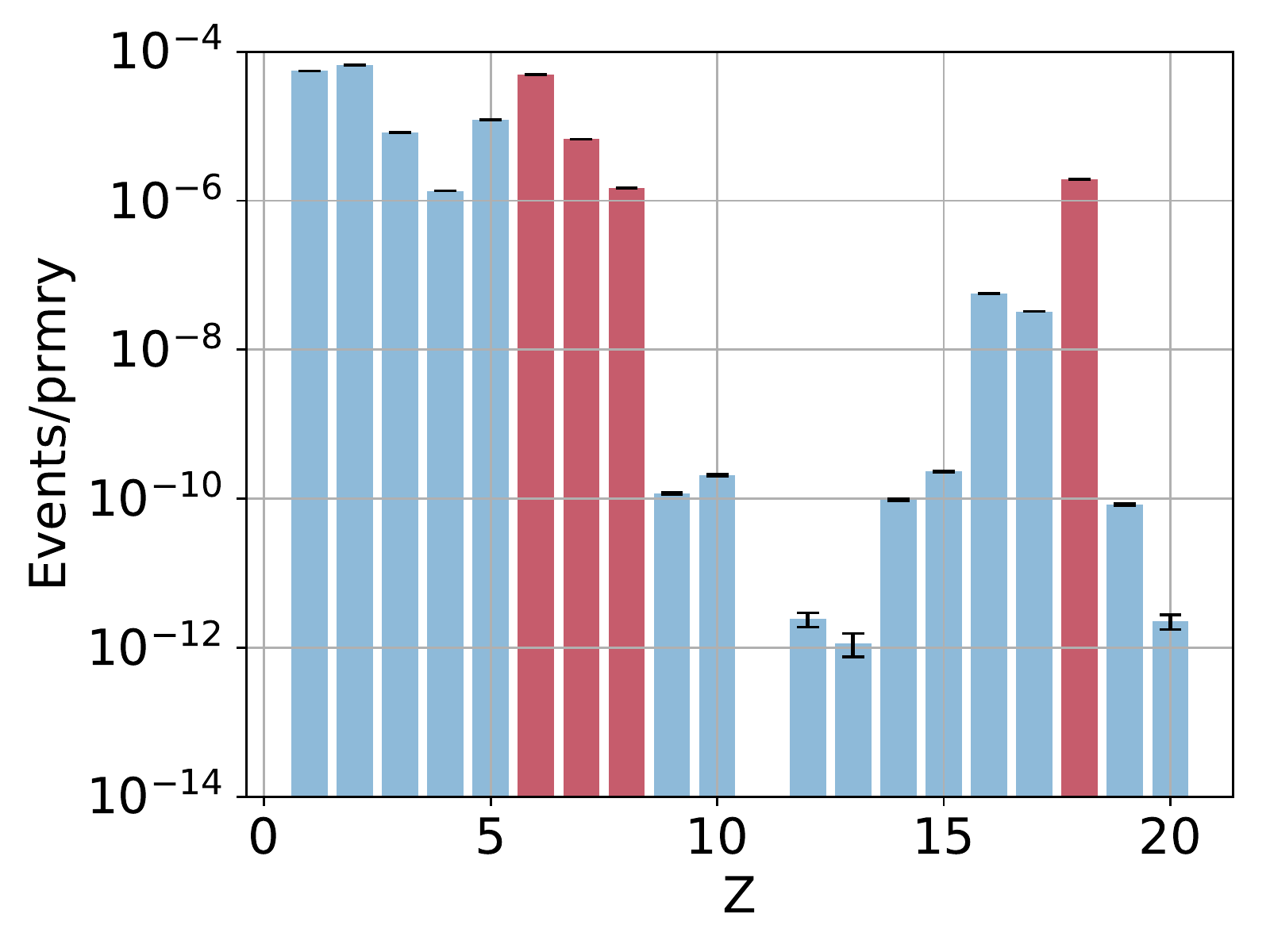}
 \noindent\includegraphics[width=.5\textwidth]{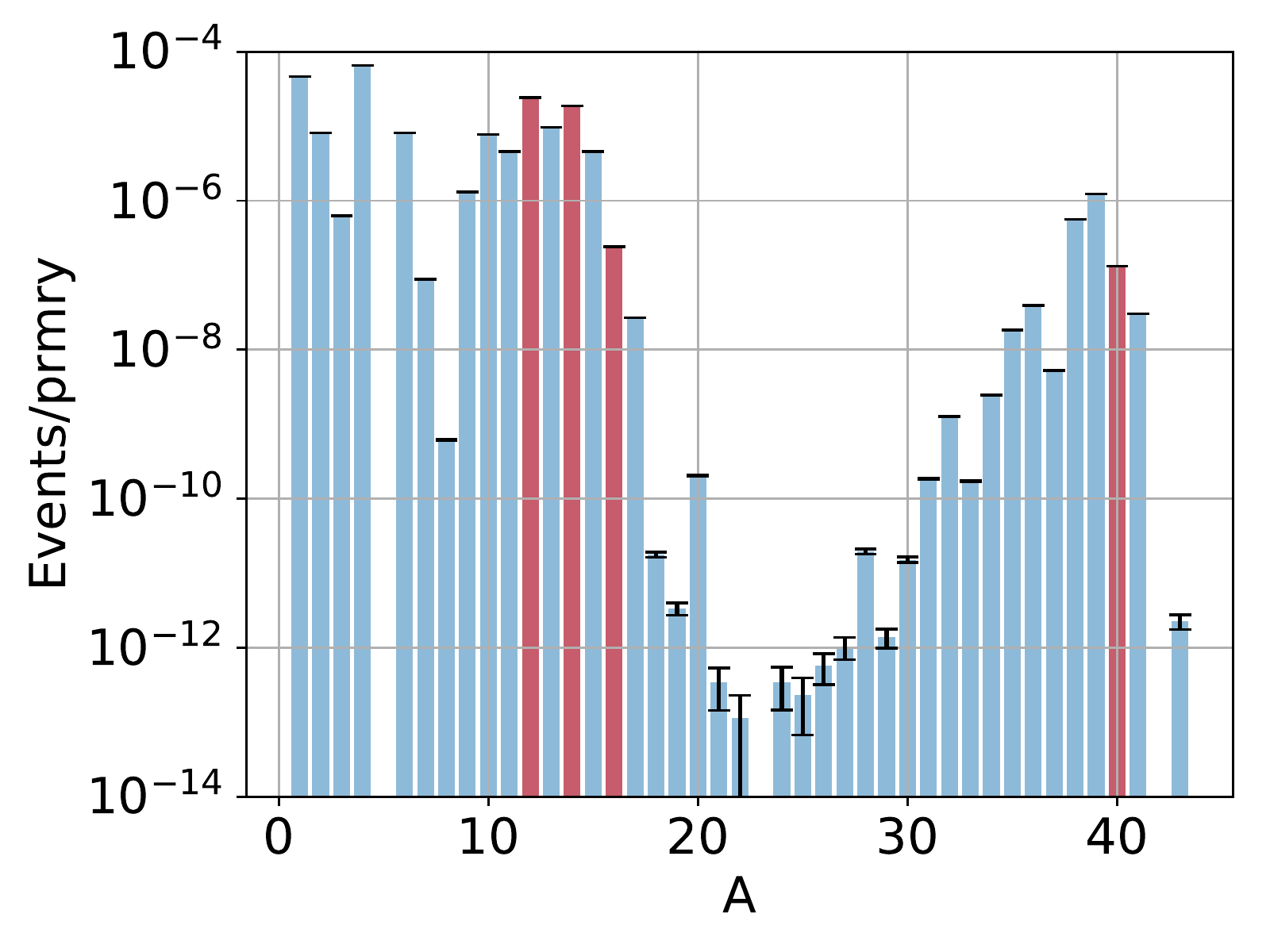}
\caption{Top: Residual nuclei in the medium after the TGF simulation. We observe, in color, the abundance of isotopes in a $A$ vs $Z$ plot, in events per primary. Bottom left: Residual nuclei abundance gathered by their $Z$ number. Bottom right: Residual nuclei collected by their $A$ number. In red, the $Z$ and $A$ of the stable $^{12}C$, $^{14}N$, $^{16}O$ and $^{40}Ar$ nuclei, which represents the major components of the air mixture.}
\label{fig:Resnuc}
\end{figure}

The full production of isotopes from the aforementioned $\gamma$ simulation is shown in Figs.~\ref{fig:Resnuc}. The simulation predicts a total yield of $1.18\cdot 10^{-4}$ nuclei per primary electron, most of them concentrated around the stable $C$, $O$, $N$ and $Ar$ isotopes of air. There is a significant production of non-stable nuclei, from $H$ to $Ca$. Regarding heavy nuclei, there is a production peak at Argon, which is fragmented and transformed into lighter nuclei up to $Z=11$, and even heavier elements such as $K$ and $Ca$, although their abundance is residual. Most of the Argon is, though, fragmented in $S$ and $Cl$ isotopes via spallation reactions with neutrons and protons. 

However, the abundance of the latter heavy isotopes is several orders of magnitude lower than those of lighter isotopes (with $Z<8$), which, due to ion fragmentation and spallation reactions, are produced in larger amounts. Some interesting light isotopes are those of beryllium, mostly $^7Be$ and $^{10}Be$, formed by spallation of oxygen and nitrogen in the atmosphere, for example via the reaction $^{16}O+n\to 4p+3n+^{10}Be$. Assuming all of this beryllium comes from cosmic-rays, the production ratio between these two isotopes can be used to study diffusive transport processes in the stratosphere and stratosphere-troposphere exchange~\cite{BOERING2015450} due to their different half lives or, from the $^{10}Be$ alone, to study past solar activity cycles. The production rate of $^{10}Be$ in the atmosphere caused by cosmic rays is around $1.8\cdot 10^{-2}$ atoms per cm$^{-2}$ per second, and the  $^7Be/^{10}Be$ ratio ranges from 1.8 in the Troposphere to 0.13 in the Stratosphere~\cite{TUREKIAN2003261}. In our simulation, we find a production rate per primary electron of $2.25(2)\cdot 10^{-8}$ $^7Be$ atoms and $1.15(5)\cdot 10^{-9}$ $^{10}Be$ atoms, which is a ratio of around $20$. 

RHESSI satellite continuously observed TGF phenomena until its decommission in 2018, showing a higher occurrence of TGFs in tropics, because in tropics the associated intra-cloud lightnings are at higher altitude~\cite{Smith1085}. RHESSI detected a rate of 10 to 20 TGF per month, which corresponds to $\sim 50$ per day globally, but a higher rate of even two orders of magnitude higher are not discarded~\cite{Smith1085}, assuming part of the global events could be undetected because they are beamed.
Let's assume a production of $1$ TGF events per second in all Earth's atmosphere~\cite{10.1029/2006JD008340}. Assuming $10^{18\pm2}$ $e^-$ per TGF, that gives us a production rate of $1.15\cdot 10^{-10\pm2}$ $^{10}Be$ atoms per cm$^{-2}$ per second, much lower than the cosmic-ray production ratio. Thus, under these assumptions, the beryllium production in thunderstorms can be neglected. Anyway, the concentration can locally peak in regions with more density of TGF events, considering that all the beryllium produced in a lightning discharge can be dissolved in rainfall and absorbed by the soil.

\begin{figure}
 \noindent\includegraphics[width=.55\textwidth]{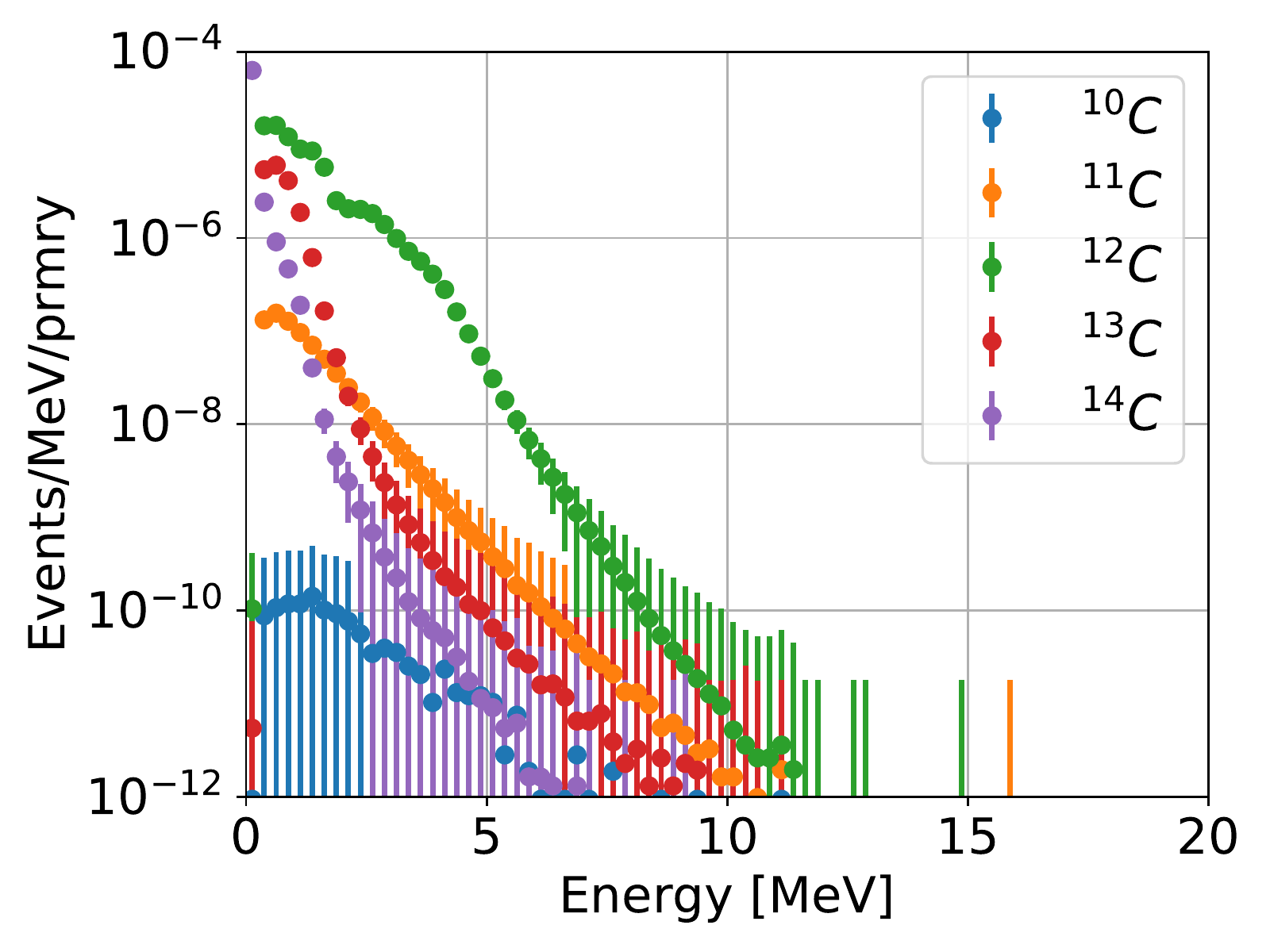}
 \noindent\includegraphics[width=.55\textwidth]{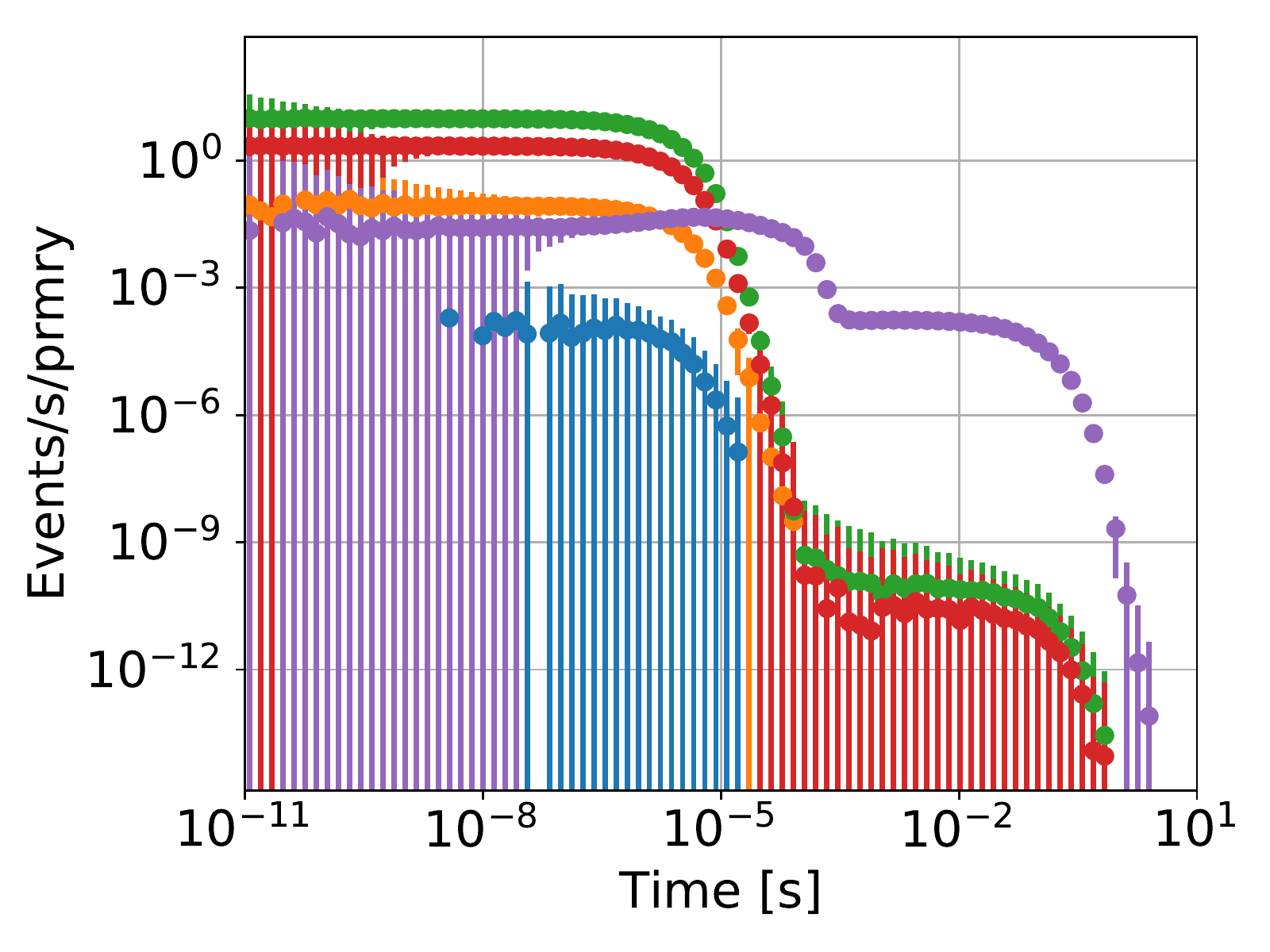}
 \caption{Left: Energy distribution of most common residual Carbon isotopes. The isotopes are mostly created with low energies, below 10 MeV. Right: Time of creation of Carbon isotopes. All isotopes, except $^{14}C$, are created within few $\mu$s after the TGF, with a small enhancement at the ms scale due to TGF afterglows. For $^{14}C$, there is a significant production at the afterglow, due to the spallation of nitrogen atoms with thermalized neutrons.}
\label{fig:Isotopeage}
\end{figure}

\begin{figure}
 \noindent\includegraphics[width=.55\textwidth]{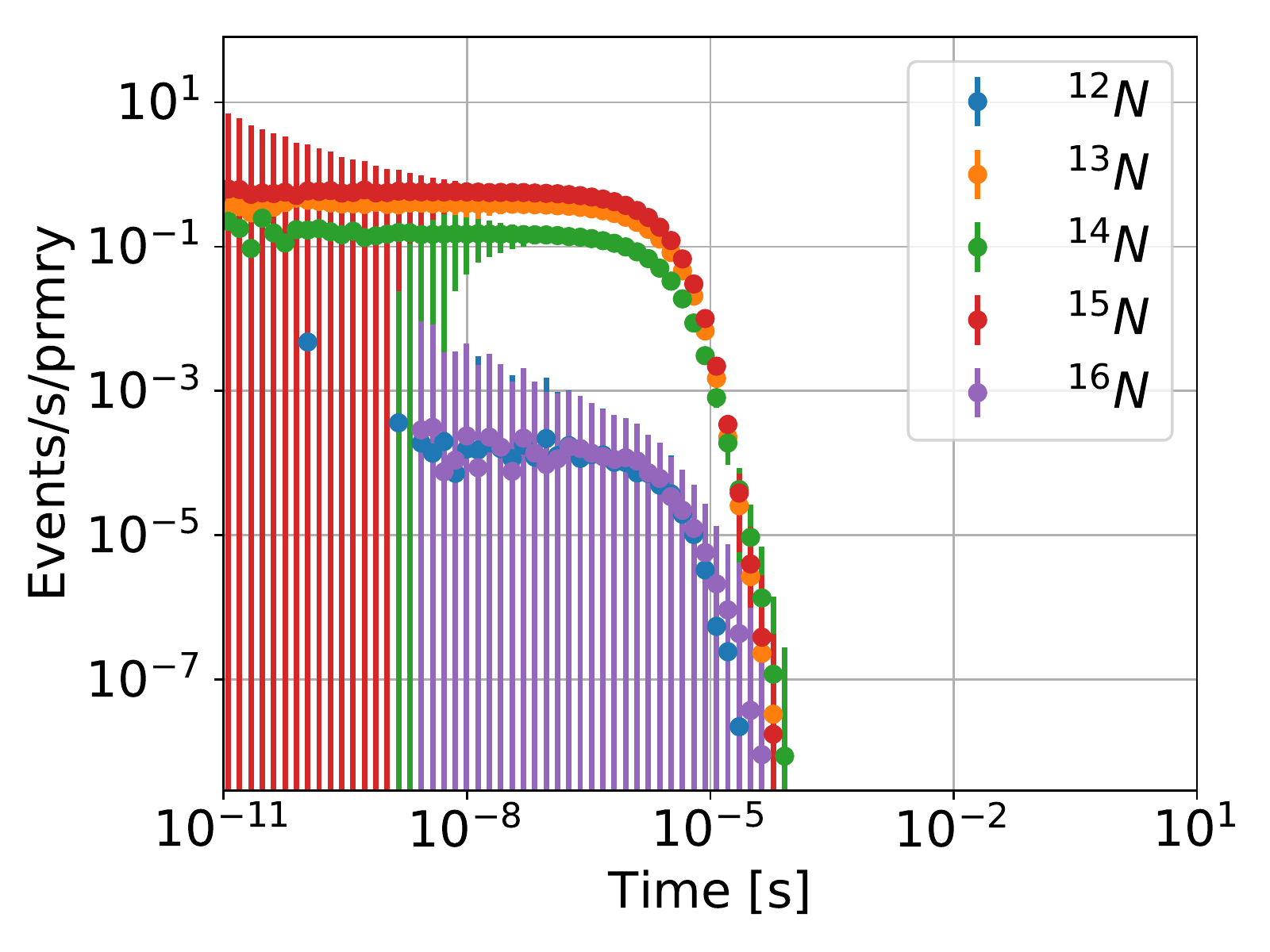}
 \noindent\includegraphics[width=.55\textwidth]{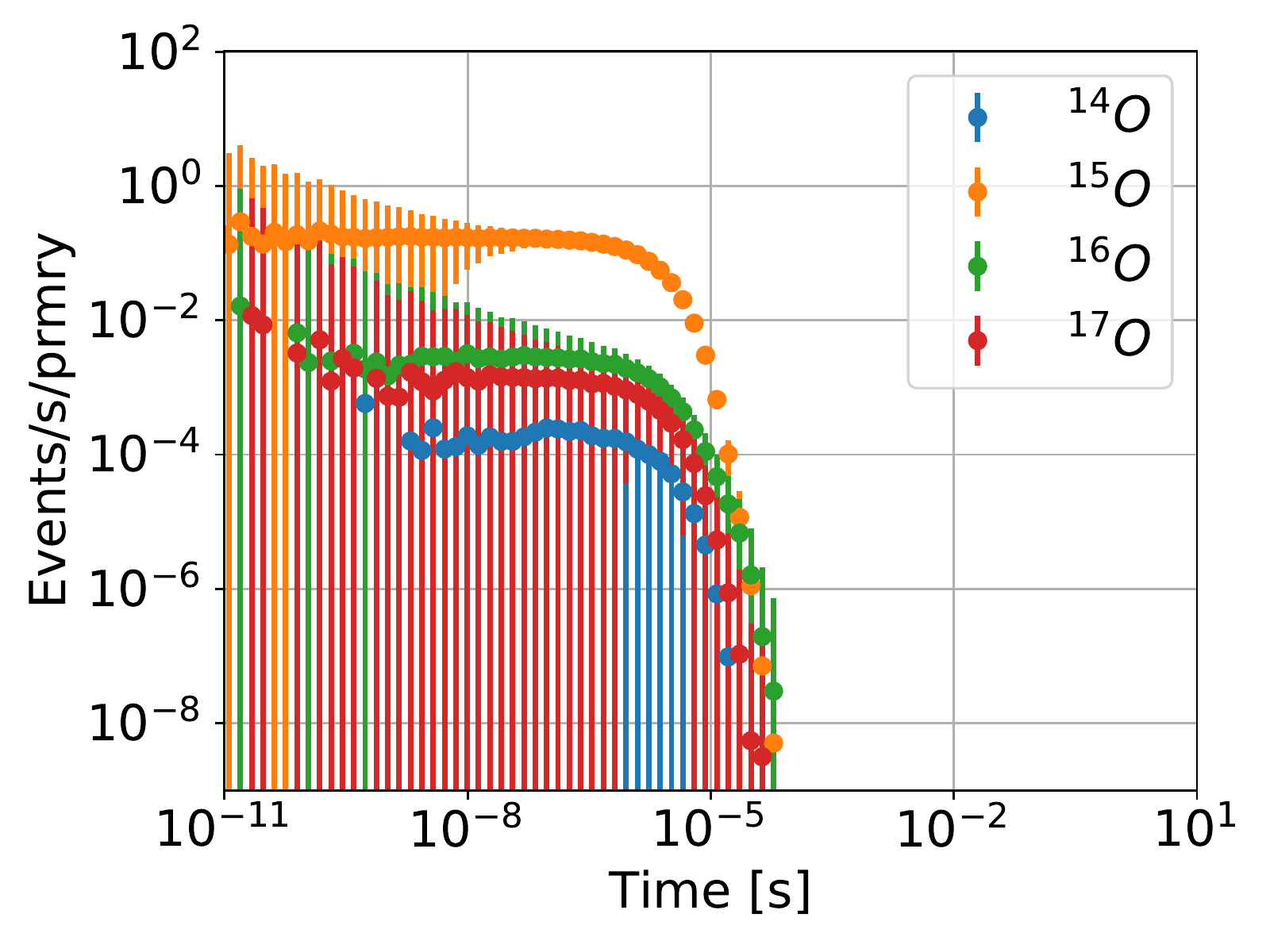}
 \caption{Time of creation of Nitrogen (left) and Oxygen (right) isotopes. Due to the limited statistics of the simulation, only the production associated with the main gamma burst appears. The trend would be similar to, for example, $^{12}C$ in Fig.~\ref{fig:Isotopeage}(right), where the second peak is around 6 orders of magnitude lower in intensity.}
\label{fig:Isotopeage2}
\end{figure}

Because of their relevance, we focus now on isotopes of nitrogen, oxygen and carbon. We can see the time profile at creation point of $N$, $O$ and $C$ isotopes in Figs.~\ref{fig:Isotopeage} and~\ref{fig:Isotopeage2}, and the energy profile of $C$ isotopes.
Most of these isotopes are created within few nanoseconds after the initial avalanche. There are some exceptions, such as $^{14}C$, which has a second production plateau in the $ms$ scale, due to the thermalization of fast neutrons created in photo-nuclear reactions with $^{14}N$ and $^{16}O$ nuclei within the atmosphere.
Basically, all the carbon isotopes are generated after inelastic interactions (with $\gamma$, light nuclei or protons), that is why they follow the same trend as the $\gamma$ time distribution in Fig.~\ref{fig:gammas}. However, the main production of $^{14}C$ comes from the scattering of $^{14}N$ with low-energy neutrons, via the reaction

\begin{equation}
    n({\rm slow})+^{14}N \to ^{14}C+^1H
\end{equation}

The additional time of flight of thermal neutrons shifts the abundance peaks to larger time values.
This is a different mechanism that the one that, for example, generates tritium, which involves the scattering of fast neutrons with stable nitrogen atoms:

\begin{equation}
    n(fast) + ^{14}N \to ^{12}C+^3H
\end{equation}

The higher the energy of the neutron, the larger the fraction of the compound nucleus ejected. TGF's are not the only source of $^{14}C$ in the atmosphere. The dominant reaction is, though, the interaction of cosmic rays with air. That production of $^{14}C$ peaks up to approximately 22000 atoms per s and m$^2$ of the earth's surface~\cite{Babich_2019,CHOPPIN2013373}. In this case, the rate is only $\sim 10^{-5}$ atoms per primary, which, estimating a mean intensity of $10^{18\pm2}$ $e^-$ per TGF, gives us  a rate of $10^{13\pm2}$ atoms of $^{14}$ C. Assuming a rate of $\sim 1$ TGF per second in the whole Earth's atmosphere, we have a ratio of $\sim 2\cdot 10^{-2\pm 2}$ atoms per s per m$^2$ of the Earth's surface. Taking the upper limit of $10^{20}$ $e^-$, the $^{14}C$ production is around a $0.01\%$ of the production from cosmic rays, but higher concentrations are not discarded due to the non-uniform distribution of TGFs at the globe~\cite{10.1029/2006JD008340,doi:10.1002/2017GL075131}. For example, the concentration would peak to $10^{7\pm2}$ ${^{14}C}$ atoms per s per m$^2$ for areas of 1 km$^2$.

Due to their importance, we show in Fig.~\ref{fig:Isotopeage2} the time profile of Nitrogen and Oxygen isotopes. Their energy distribution is similar to that of the Carbon isotopes (Fig.~\ref{fig:Isotopeage}(left)), staying below 10 MeV. 
The isotope production in the simulation approximate follows the well-known Boltzmann distribution. Indeed, the occupation number of atoms per energy is proportional to:

\begin{equation}
    \frac{d^3N}{dE} \propto E\sqrt{E^2-m^2} e^{-\frac{E}{T}}
\end{equation}
where $E(p)=\sqrt{p^2+m^2}$ is the energy of the particle and $T$ the temperature. The characteristic temperature $T$ of the process depends on the isotope. It ranges between 3 MeV for light ions ($Z<2$) to 0.5 MeV for Carbon, following the approximate relation with the atomic mass number of $T=(6\pm2 MeV)\cdot A^{-1}$.

Besides the stable $^{14}N$, the most produced isotopes of Nitrogen are $^{13}N$ and $^{15}N$. Regarding the Oxygen, there is a clear preference for $^{15}O$, between one and two orders of magnitude larger than the rest of Oxygen isotopes. This is an interesting result because $^{15}O$ is a relatively long-lived $\beta^+$ radio-isotope of the Oxygen, being its half-life of around 122 seconds. Their production can be measured by ground-based facilities (for lower TGFs) as a $0.511$ MeV  $\gamma$ peak long after the TGF has been produced~\cite{Enoto:2017lpx}.
Additionally, there are other $\beta^+$ or electron-capture (EC) emitters with sufficiently large production which could be detected. Their time decay patterns covers wide time scales. However, long-lived isotopes, such as $^7$Be or $^{11}$C, can be dispersed by air currents and diffused over larger areas before they decay, making them difficult to measure and identify.

 \begin{table}
 \caption{\label{tab:isotopes} Table of the production abundances (per primary $e^-$) of $\beta^+$ emitters in the TGF event simulated with FLUKA.}
 \centering
 \begin{tabular}{l l l l l}
 \hline
  Isotope  & Production & Mechanism & Half-life & Daughter \\
 \hline
$^7$Be    & $\left(2.25\pm 0.02\right)\cdot 10^{-8}$ & EC & 53.33(6) d & $^7$Li \\  
$^8$B     & $\left(2.3\pm0.2\right)\cdot 10^{-10}$ & EC,$\beta^+$ & 770(3) ms  &  $^{8}$Be\\
$^{10}$C  & $\left(8.5\pm1.4\right)\cdot 10^{-11}$ & $\beta^+$ &19.3009(17) s  &  $^{10}$B\\
$^{11}$C  & $\left(1.18\pm0.01\right)\cdot 10^{-7}$ & $\beta^+$ &20.364(14) min  &  $^{11}$B\\
$^{12}$N  & $\left(1.8\pm0.2\right)\cdot 10^{-10}$ & $\beta^+$ & 11.000(16) ms &  $^{12}$C\\
$^{13}$N  & $\left(9.18\pm0.02\right)\cdot 10^{-7}$ & $\beta^+$ &9.965(4) min  & $^{13}$C\\
$^{14}$O  & $\left(3.9\pm 0.3\right)\cdot 10^{-10}$ & $\beta^+$ & 70.620(13) s & $^{14}$N\\
$^{15}$O  & $\left(6.54\pm 0.01\right)\cdot 10^{-7}$ & $\beta^+$ & 122.24(16) s & $^{15}$N\\
 \hline
 \end{tabular}
 \end{table}

\section{Conclusions}

In this work we have analyzed the secondaries and isotope creations after a RREA event at three different heights. The production of both light and heavy nuclei, specially those with measurable decay modes (such as medium and long-lived $\beta^+$ emitters) can be of interest to characterize the gamma burst and study their properties in detail.

For that purpose we have employed the widely-used Monte Carlo particle transport code FLUKA to simulate a full relativistic runaway electron avalanche similar to those that generates TGF events. The ensuing EM shower, and specially the high-energy $\gamma$ flux, are found to create a rich production of non-stable nuclei due to photo-nuclear interactions. The time profile of carbon, oxygen and nitrogen isotopes are shown in detail, showing two timescales of production, one at the nanosecond scale (where the main gamma flash is produced, associated with photo-nuclear reactions and spallation with fast neutrons) and another one in the millisecond scale (the so-called TGF afterglow, with smaller intensity, associated mainly with thermalized neutrons). Special attention for $^{14}C$ isotope, which has a large production rate compared to the rest of  carbon isotopes due to the scattering of nitrogen atoms with slow neutrons produced in photo-nuclear interactions after the gamma-ray flash.

The initial height of the TGF is found to have a negligible effect in which respects the secondary spectrum, although it has an influence on the spread of the created particles due to the decrease of density with height in the atmosphere, which increases the atmospheric mean free path. The initial height of the TGF would, nevertheless, have an effect on the detection capability for satellite and ground-based facilities.

The atmospheric photo-nuclear reactions triggered by lightnings provide a previously unexplored channel for generating isotopes of carbon, nitrogen and oxygen naturally on Earth, opening new detection mechanisms that can help us to characterize this high-energy atmospheric events. 
The enhancement in isotopes abundances, specially unstable nuclei decaying through $\beta^+$ mechanisms can produce glows in the seconds to minutes scales after the TGF, giving an additional tool to evaluate the high-energy event. We have estimated the production of some $\beta^+$ emitters, showing that the most promising ones are $^{15}$O, $^{13}$N and $^{11}$C, as they are the most abundant, all of them decaying in the minutes scale (see Tab.~\ref{tab:isotopes}).

Finally, Fig.~\ref{fig:Resnuc} showed the family of residual nuclei created after a TGF event. Some of them are long-lived radionuclides which are used as tracers for archaeometry, such as tritium, $^{14}C$ or $^{10}Be$. The production rate for the latter radionuclides are $\sim 10^{-5}$ atoms per primary for $^{14}C$, $\sim 10^{-7}$ for $^3H$ and $\sim 10^{-9}$ for $^{10}Be$. Considering a population of $10^{20}$ $e^-$ per TGF and a rate of $\sim 1$ TGF per second in the whole Earth's atmosphere, the production rate would be $\sim 2$ atoms m$^{-2}$ s$^{-1}$ for $^{14}C$, $\sim 0.02$ atoms m$^{-2}$ s$^{-1}$ for $^3H$ and $2\cdot 10^{-4}$ atoms m$^{-2}$ s$^{-1}$ for $^{10}Be$. Compared to the main production mechanism of such radionuclides, cosmic rays, this production corresponds to a fraction of $10^{-2}\%$ of the global $^{14}C$ production, $10^{-3}\%$ of the $^3H$ production and $10^{-4}\%$ of the $^{10}Be$~\cite{CHOPPIN2013373}. That is, a priori, a marginal productions unless the TGF events are concentrated on severe thunderstorm activity areas, in which case the latter concentrations may locally peak up~\cite{10.1029/2006JD008340,doi:10.1002/2017GL075131}.

\section*{Acknowledgements}
This work has been partially funded by Spanish Ministerio de Econom\'ia, Industria y Competitividad under 
contract FPA2016-77177-C2-2-P and by the EU STRONG-2020 project under the program H2020-INFRAIA-2018-1, grant agreement no.824093. Data employed in this work were provided by the Monte Carlo code FLUKA, available at www.fluka.org and https://fluka.cern.

\bibliography{IsotopeTGF}

\end{document}